\documentclass{jpsj2}
\newcommand{\bq}{\boldsymbol{q}}
\newcommand{\bqp}{\boldsymbol{q}^{\prime}}
\newcommand{\bqpp}{\boldsymbol{q}^{\prime\prime}}
\newcommand{\bqppp}{\boldsymbol{q}^{\prime\prime\prime}}
\newcommand{\bm}{\boldsymbol{m}}
\newcommand{\bK}{\boldsymbol{K}}
\newcommand{\bR}{\boldsymbol{R}}
\newcommand{\bQ}{\boldsymbol{Q}}
\newcommand{\be}{\boldsymbol{e}}

\title{Multiple Helical Spin Density Waves and Magnetic Skyrmions in Itinerant Electron System}

\author{Yoshiro Kakehashi\thanks{yok@sci.u-ryukyu.ac.jp, to be published in J. Phys. Soc. Jpn. (2018).}, Daiki Koja,  Todgerel Olonbayar, and Hideyori Miyagi}
\inst{Department of Physics and Earth Sciences, Faculty of Science,\\
University of the Ryukyus, \\
1 Senbaru, Nishihara, Okinawa, 903-0213, Japan} %\\

\abst{
Magnetic structures and their stability of the multiple helical spin density waves (MHSDW) and related magnetic skyrmions in itinerant electron system have been investigated on the basis of both the phenomenological and microscopic theories.  It is shown by using the Ginzburg-Landau (GL) theory and the Application Visualization System (AVS) that the magnetic skyrmion structures emerge as a MHSDW and can be stabilized in itinerant electron system on the fcc lattice even if there is no Dzyaloshinskii-Moriya interaction.  Moreover, by using the generalized Hartree-Fock theory for the Hubbard model, the antiferromagnetic skyrmion structure found in the GL theory is shown to be stabilized on the fcc lattice in the vicinity of the half-filled electron number.
}

\kword{  itinerant magnetism, multiple helical spin density waves, magnetic skyrmions, 
antiferromagnetic skyrmions, Ginzburg-Landau theory, Hubbard model, 
generalized Hartree-Fock approximation }

\begin{document}
\maketitle

\section{Introduction}

Recently, a new type of magnetic structure called magnetic skyrmions has been found in B20 transition metal compounds and their alloys such as MnSi~\cite{muhl09}, Fe${}_{1-x}$Co${}_{x}$Si~\cite{munz10,yu10}, and FeGe~\cite{yu11}.
The magnetic skyrmions are a vortex-like magnetic structure with particle properties~\cite{bog01,ross06,heinz11}.
Due to the lack of inversion symmetry in the B20 crystal structure, the formation of magnetic skyrmions is usually well explained by a localized model with competition between the ferromagnetic exchange interaction and the Dzyaloshinskii-Moriya (DM) interaction~\cite{dzya58,mori60} under magnetic field~\cite{bak80,yi09}.  More recently, it has been found that competing magnetic interactions on the frustrated lattice can also stabilize the magnetic skyrmion structures.   Okubo, Chung, and Kawamura have shown on the basis of the Monte-Carlo simulation that the Heisenberg model without the DM interaction can stabilize the magnetic skyrmion structure on the triangular lattice due to competing magnetic interactions~\cite{okubo12}.
Lin and Hayami found using the Ginzburg-Landau type of theory that competing interactions with easy-axis anisotropy can stabilize the skyrmions~\cite{lin16}.  Quite recently, Ozawa and coauthors reported that the Kondo lattice model without DM interaction can also produce the magnetic skyrmion structure via the Fermi surface effect of conduction bands~\cite{ozawa17, hayami17}.

Although the mechanisms mentioned above have been well established, most of the theories rely on the Heisenberg model or its continuum version.  Even the Kondo lattice model is based on the localized picture that the magnetic $d$ or $f$ electrons are localized and are coupled with conduction bands via exchange interaction, and thus cannot be applicable to the typical itinerant electron system such as transition metals and their alloys, where  magnetic $d$ electrons themselves are delocalized and form clear bands with band width $W   \sim U$, $U$ being the Coulomb interaction strength. 
The itinerant electron system is wellknown to be described by the Hubbard model~\cite{hub63,hub64,gutz63,gutz64,kana63}.  
The magnetic skyrmion structure in the itinerant magnetism and the role of the underlying electronic structure therefore have not yet been understood, though some similarities between the Kondo and Hubbard models have been reported for the formation of multiple spin density waves~\cite{mart08,akagi10}.

We studied previously various magnetic structures in itinerant electron system on the basis  of the Ginzburg-Landau (GL) theory~\cite{uchida06} which allows for the amplitude change of local magnetic moments at the ground state, and developed the microscopic theory of molecular spin dynamics method~\cite{kake98,kimu00,uchida16} to clarify the magnetic structures on the basis of the electronic structure of the system.  In particular, we pointed out that $\gamma$-Fe is possibly a multiple helical spin density waves (MHSDW) with wave vector (0.1,0,1)$2\pi/a$, $a$ being a lattice parameter~\cite{uchida06}.
Since the magnetic skyrmions are a MHSDW, it is worthwhile to investigate the magnetic skyrmion structure from the viewpoint of the GL theory in more details and to clarify various magnetic structures of the MHSDW in relation to the skyrmions.

In this paper, we investigate the stability of the MHSDW and the magnetic skyrmions in  itinerant electron system on the fcc lattice on the basis of the phenomenological GL theory without the DM interaction, and clarify their magnetic structures in the real space using the AVS (Application Visualization System) image analysis.  Moreover we present some numerical results of calculations for the existence of  the magnetic skyrmion structure in the Hubbard model.  This is the first attempt to bridge the gap between the standard framework of itinerant electron magnetism~\cite{moriya85,kake13} and  the skyrmion physics.

We will demonstrate using the AVS image approach and the GL theory that the magnetic skyrmions appear indeed in itinerant electron system as a MHSDW, and can be stabilized by a competition between long-range ferro- and antiferro-magnetic interactions even if there is no DM interaction.
Moreover, we show that the antiferromagnetic (AF) base skyrmions appear when the wave vectors $\bQ$ approach to the antiferromagnetic ones on the Brillouin zone boundary.  We also demonstrate that the AF-base skyrmions can be stabilized in the vicinity of the half-filled electron number in the Hubbard model on the fcc lattice, using  the generalized Hartree-Fock (GHF) theory.

The paper is organized as follows.  In the following section, we briefly review the GL theory for the MHSDW on the fcc lattice and show that the MHSDW are possible when the couplings between two helical states are not so large as compared with the fourth-order coupling of the individual mode.  In \S 3, we clarify the real-space magnetic structures for the MHSDW using the AVS image method.
We find that the triple $\bQ$ (3Q) MHSDW include the vortex-type skyrmions, the hedgehog-type skyrmions, the antiferromagnetic skyrmions, as well as the $\gamma$-Fe type MHSDW.
In \S 4, we present some numerical results of calculations for the single-band Hubbard model on the fcc lattice, which are based on the GHF approximation, and demonstrate that the AF-base skyrmions are stabilized in the vicinity of half-filling.  In the last section, we summalize the results and discuss some problems which remain unsolved in the present work.

\section{Ginzburg-Landau Theory for Multiple Helical Spin Density Waves}

Ginzburg-Landau phenomenological theory for the multiple helical spin density waves (MHSDW) in itinerant electron system has been presented by Uchida and Kakehashi~\cite{uchida06}.  We describe here the essence of the theory as an introduction to the next section.  We consider a cubic system with an atom per unit cell and a local magnetic moment $\langle \bm_{l} \rangle$ on site $l$, and expand the free energy with respect to the local moments up to the fourth order.
In the isotropic system without external magnetic field, the free energy per site is expressed in the Fourier representation as follows.
\begin{equation}
f = \sum_{\bq} A(\bq)|\bm(\bq)|^2 
+\sum_{\bK}\sum_{\bq,\bqp,\bqpp,\bqppp} B(\bq,\bqp,\bqpp,\bqppp)
\{\bm(\bq)\cdot\bm(\bqp)\}\{\bm(\bqpp)\cdot\bm(\bqppp)\} .
\label{fourierf}
\end{equation}
Here the sum $\sum_{\bq}$ ($\sum_{\bK}$) is taken over the Brillouin zone (the reciprocal lattice points such that $\bK = \bq + \bqp + \bqpp + \bqppp$). 
$A(\bq)$ and $B(\bq,\bqp,\bqpp,\bqppp)$ are expansion coefficients of the second- and fourth-order terms in the Fourier representation. $\bm(\bq)$ is a Fourier representation of magnetic moments $\{ \langle \bm_{l} \rangle \}$, defined by 
$\bm(\bq) = \sum_{l} \langle \bm_{l} \rangle \exp (-i\bq \cdot \bR_{l}) / N$. 
$N$ is the number of lattice sites, and $\bR_{l}$ is the position vector on site $l$.
It should be noted that the free energy (\ref{fourierf}) works best for the weak itinerant magnets because of the lowest-order expansion with respect to the magnetic moments.

The triple $\bQ$ multiple helical spin density waves (3Q-MHSDW) on the cubic lattice is expressed as
\begin{align}
\langle \bm_{l} \rangle &= m_{1} \big( \be_{2} \cos (\bQ_{1} \cdot \bR_{l} + \alpha_{1}) +
\be_{3} \sin (\bQ_{1} \cdot \bR_{l} + \alpha_{1}) \big)  \nonumber \\
&+
m_{2} \big( \be_{3} \cos (\bQ_{2} \cdot \bR_{l} + \alpha_{2}) +
\be_{1} \sin (\bQ_{2} \cdot \bR_{l} + \alpha_{2}) \big) \nonumber \\
&+
m_{3} \big( \be_{1} \cos (\bQ_{3} \cdot \bR_{l} + \alpha_{3}) +
\be_{2} \sin (\bQ_{3} \cdot \bR_{l} + \alpha_{3}) \big) .
\label{3qm}
\end{align}
Here $m_{i} \ (i = 1, 2, 3)$ is an amplitude of each helical magnetic moment. 
$\be_{1}, \be_{2}, \be_{3}$ denote the unit vectors along the $x$, $y$, and $z$ axes, respectively.
$\bQ_{i}$ and $\alpha_{i}$ are a wave vector and a phase factor of each helical SDW $i$.

Equation (\ref{3qm}) is expressed in the Fourier representation as follows.
\begin{align}
\langle \bm_{l} \rangle &= \sum_{i=1}^{3} \big( \bm(\bQ_{i}) {\rm e}^{i\bQ_{i}\cdot \bR_{l}}
+ \bm(\bQ_{i})^{\ast} {\rm e}^{-i\bQ_{i}\cdot \bR_{l}} \big) .
\label{fourierm}
\end{align}
Here $\bm(\bQ_{1}) = (0, m(\bQ_{1}), -i m(\bQ_{1}))$,   
$\bm(\bQ_{2}) = (-i m(\bQ_{2}), 0, m(\bQ_{2}))$, 
$\bm(\bQ_{3}) = (m(\bQ_{3}), -i m(\bQ_{3}), 0)$, and 
$m(\bQ_{j}) = m_{j} {\rm e}^{i\alpha_{j}} /2 \ (j=1, 2, 3)$.
Note that the helical condition of each component $i$ is given by 
\begin{align}
\bm(\bQ_{i})^{2} = 0 .
\label{hcond}
\end{align}

For the 3Q-MHSDW with incommensurate wave vectors, the free energy (\ref{fourierf}) is expressed as follows.
\begin{align}
f &= \sum_{i=1}^{3} 2 \big[ A_{Q} m(\bQ_{i})^2 + 2 B_{1Q} m(\bQ_{i})^4 \big] \nonumber \\
  &+ B_{QQH} \big[ m(\bQ_{2})^2 m(\bQ_{3})^2 + m(\bQ_{3})^2 m(\bQ_{1})^2 
                        + m(\bQ_{1})^2 m(\bQ_{2})^2 \big] .
\label{3qmhsdwf}
\end{align}
Here $A_{Q}$ and $B_{1Q}$ are coefficients for individual mode, and $B_{QQH}$ is the coefficient for the couplings between two modes.  The coupling terms among three $\bQ$'s vanish in the incommensurate case because of the helical condition (\ref{hcond}). 
\begin{figure}[htbp]
\begin{center}
\includegraphics[width=8cm]{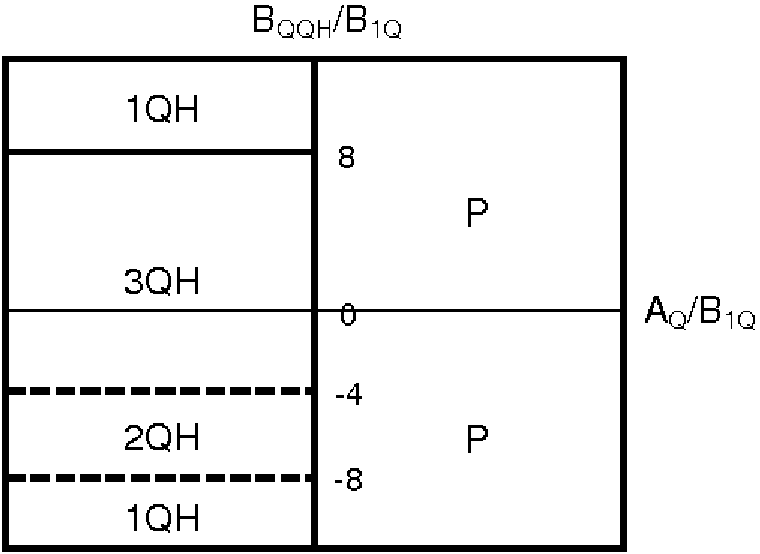}
\end{center}
\vspace{0cm}
\caption{Magnetic phase diagram of the multiple helical spin density waves (MHSDW) on the $A_{Q}/B_{1Q}$-$B_{QQH}/B_{1Q}$ plane.  1QH: single $\bQ$ helical state, 2QH: double $\bQ$ helical state, 3QH: triple $\bQ$ helical state, and P: paramagnetic state.  Dashed lines indicate the lines where the amplitudes of local magnetic moments diverge, so that the present analysis based on the fourth-order free energy is not applicable near these lines.
}
\label{fig-3qhpd}
\end{figure}

Minimizing the free energy (\ref{3qmhsdwf}) with respect to $\{ m(\bQ_{i}) \}$, we obtain the phase diagram on the $A_{Q} - B_{QQH}$ plane as shown in Fig. \ref{fig-3qhpd}.
Note that the boundary between the single $\bQ$ helical state (1QH) and the double $\bQ$  helical state (2QH), and 
that between the double $\bQ$ helical state (2QH) and the triple $\bQ$ helical state (3QH) cannot be determined by the 4th-order GL theory~\cite{uchida07} because the amplitude of local moment defined by 
$M^2 = \sum_{l} \bm_{l} \cdot \bm_{l} / N$ diverges at $B_{QQH}=-8B_{1Q}$ and $B_{QQH}=-4B_{1Q}$, respectively.
We need analysis with use of the free energy with higher-order terms to determine these boundaries. Nevertheless, we can conclude  that the 3Q-MHSDW states can be stabilized in a certain range of $-4 < B_{QQH}/B_{1Q} < 8$ when $A_{Q} < 0$ and $B_{1Q} > 0$.
We remark that the present result (Fig. \ref{fig-3qhpd}) is very similar to the $H-T$ phase diagram of frustrated Heisenberg model obtained by Okubo {\it et al.}~\cite{okubo12}, even though both approaches are quite different.
  
We discussed here only the case of MHSDW for simplicity, but the existence of the 3Q-MHSDW does not change even if we consider the other types of MSDW~\cite{uchida06,uchida07}.
Moreover we can prove that the 3QH state is always lower than the 2QH and 1QH states in energy when there is a solution of the 3QH state.  The stability of the 3QH state is caused by the increase in amplitude of local moments.  This is characteristics of the weakly correlated itinerant electron system.  In fact, the classical Heisenberg model without the DM interaction does not yield the MHSDW state at the ground state according to the Yoshimori-Kaplan theory~\cite{yoshi59,kap60} because of the constraint of the constant amplitude of local magnetic moments.

\section{AVS Image Analysis for the MHSDW}
\subsection{$\gamma$-Fe type MHSDW}
The 3QH state discussed in the last section does not yield a unique magnetic structure in the real space because the amplitude $m_{i}$ and the wave vector $\bQ_{i}$ as well as the phase factor $\alpha_{i}$ in each helical state $i$ are variable as seen in Eq. (\ref{3qm}).
Varying these parameters, a variety of magnetic structures are possible in the 3QH state.
One of the possible magnetic structures is the MHSDW that was discussed in $\gamma$-Fe.  Neutron diffraction experiments~\cite{tsuno89} for the cubic $\gamma$-Fe${}_{1-x}$Co${}_{x}$ ($x <4$) alloys precipitates in Cu show a magnetic satellite peak at the wave vector $\bQ = (q, 0, 1) 2\pi/a$ with $q=0.1$.  Here $a$ denotes the lattice constant of the fcc lattice. 
Tsunoda~\cite{tsuno89} suggested that a magnetic structure of $\gamma$-Fe is a helical SDW (1QH) in which the antiferromagnetic moments on the $y$-$z$ plane rotate along $x$ axis with the wave length $a/q$.  On the other hand, we proposed the 3QH with the wave vectors $\bQ_{1} = (q, 0, 1) 2\pi/a$, $\bQ_{2} = (1, q, 0) 2\pi/a$, $\bQ_{3} = (0, 1, q) 2\pi/a$.  We call this structure hereafter the $\gamma$-Fe type MHSDW.
\begin{figure}[htbp]
\begin{center}
\includegraphics[width=8cm]{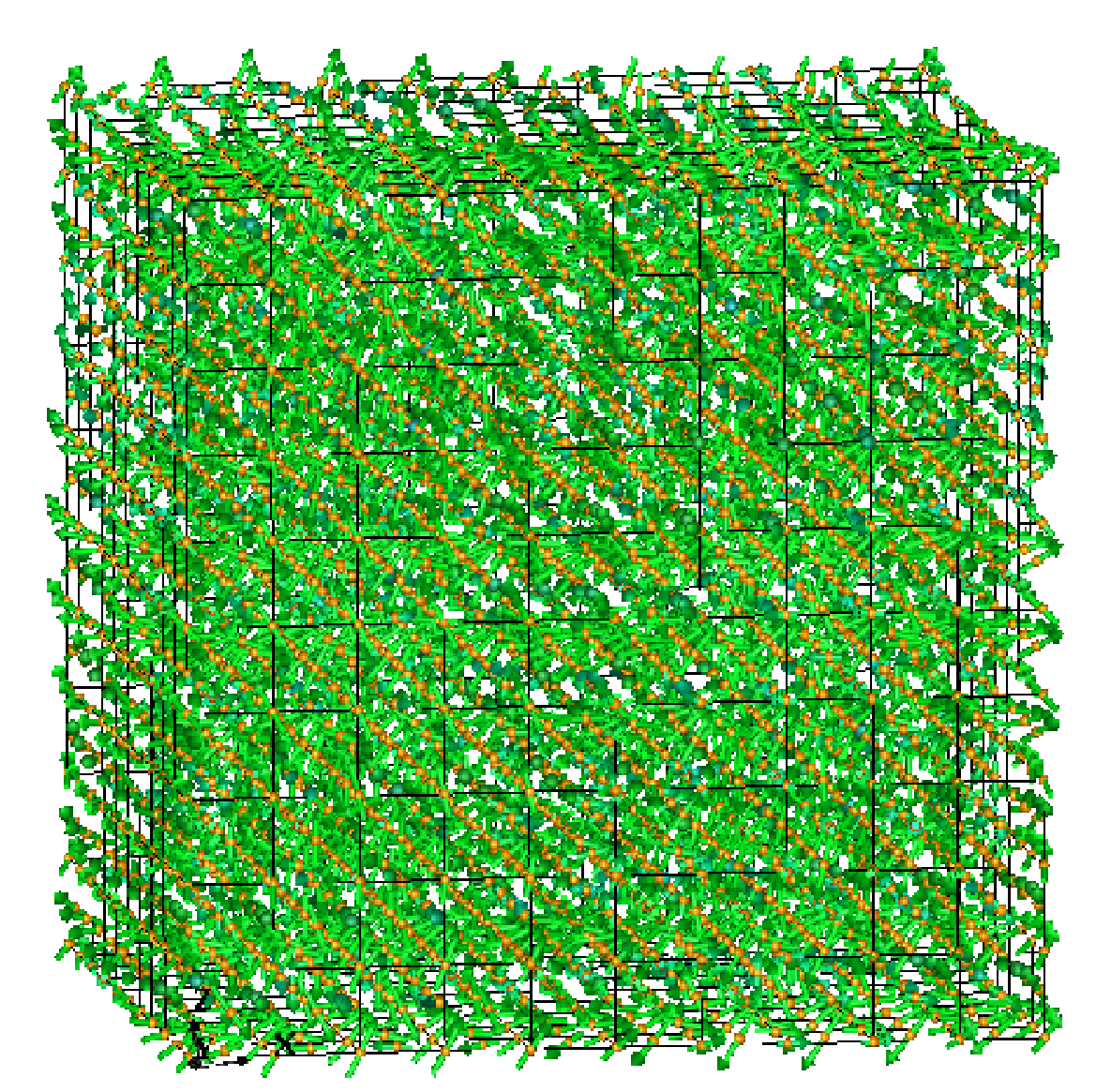}
\end{center}
\vspace{0cm}
\caption{Magnetic structure of the $\gamma$-Fe type MHSDW for $q=0.1$.
}
\label{fig-femhsdw}
\end{figure}

\begin{figure}[htbp]
\begin{center}
\includegraphics[width=8cm]{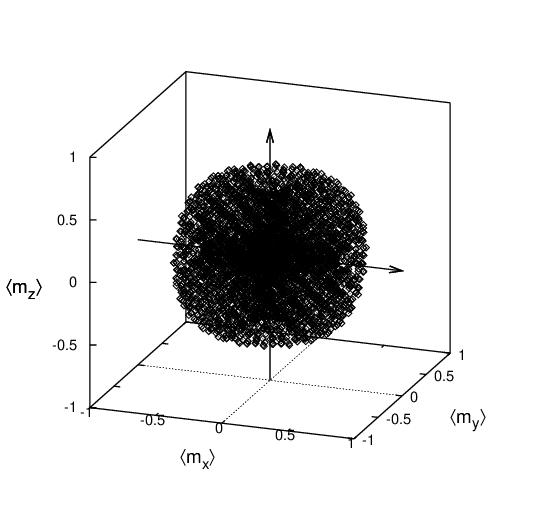}
\end{center}
\vspace{0cm}
\caption{Local moment distribution for $\gamma$-Fe type MHSDW ($q=0.1)$, which was obtained from 4000 local moments in the $10 \times 10 \times 10$ fcc unit cells.
}
\label{fig-femhsdwlm}
\end{figure}

The 3Q-MHSDW show the complex magnetic structure in general.  One cannot recognize the structure in the real space from Eq. (\ref{3qm}) or Eq. (\ref{fourierm}) with Eq. (\ref{hcond}).  Therefore we adopted the AVS image method for understanding the magnetic structures intuitively.  We have created the 3QH magnetic structure on a large cluster consisting of $10 \times 10 \times 10$ fcc unit cells with periodic boundary condition, and have drawn the magnetic structure using the AVS.  Possible magnetic structure is presented in Fig. \ref{fig-femhsdw} for $q=0.1$, $m_{1}=m_{2}=m_{3}=0.3$, and $\alpha_{1} = \alpha_{2} = \alpha_{3} = 0$.  We observe that the local magnetic moments rotate clockwise along the $x$, $y$, $z$ axes, respectively, with changing their amplitudes.  The local-moment distribution obtained from 4000 local moments in the cluster is presented in Fig. \ref{fig-femhsdwlm}.  The distribution is almost uniform in direction, but it is not on a sphere because of the change of amplitude.  

\subsection{Skyrmion-type MHSDW}

\begin{figure}[htbp]
\begin{center}
\begin{minipage}{9cm}
\begin{center}
\includegraphics[width=9cm]{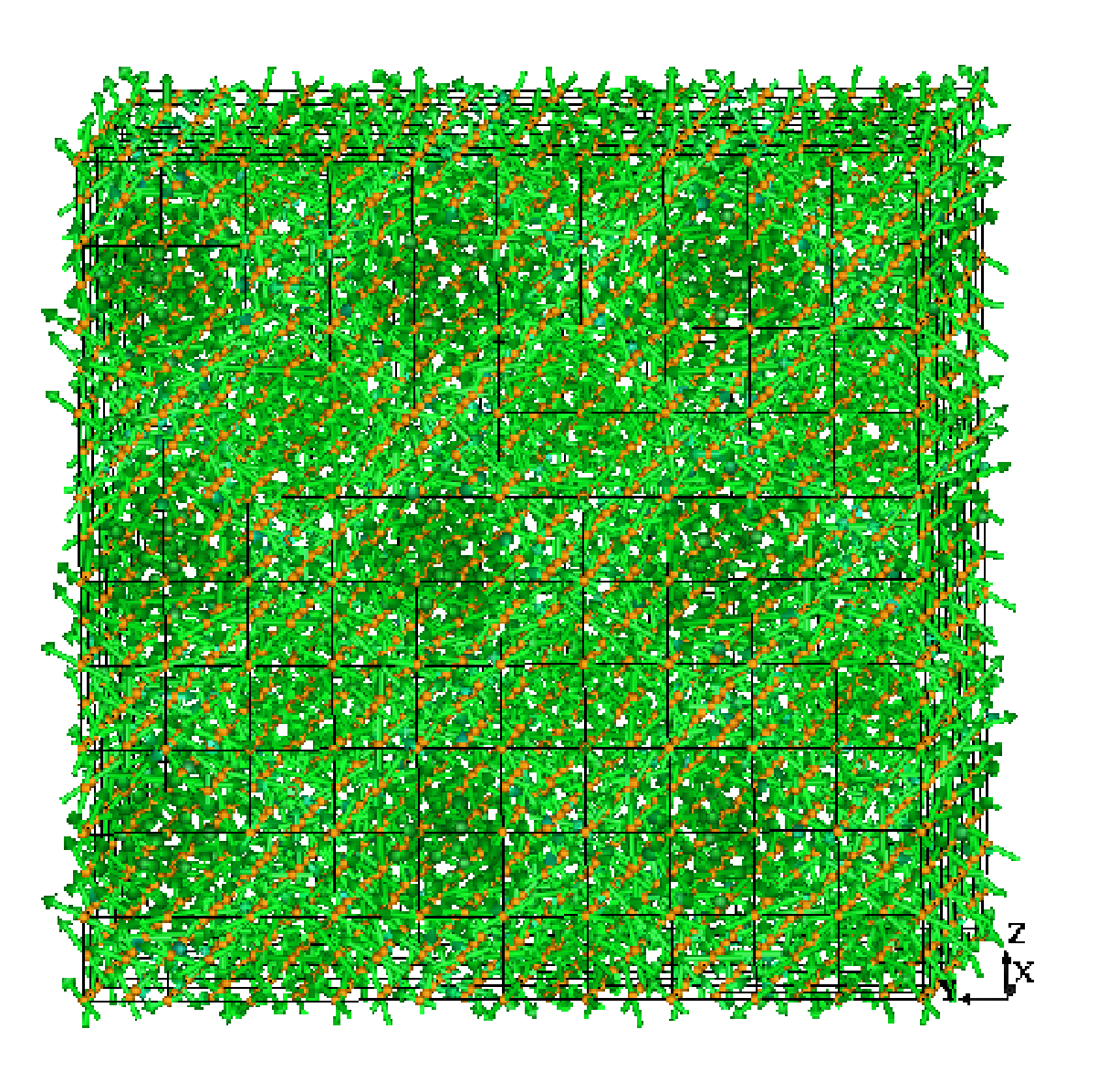} 
\vspace{-1cm} \\
(a) 
\vspace{5mm} \\
\end{center}
\end{minipage}
\hspace{0.7cm}
\begin{minipage}{5cm}
\begin{center}
\vspace*{3cm}
\includegraphics[width=5cm]{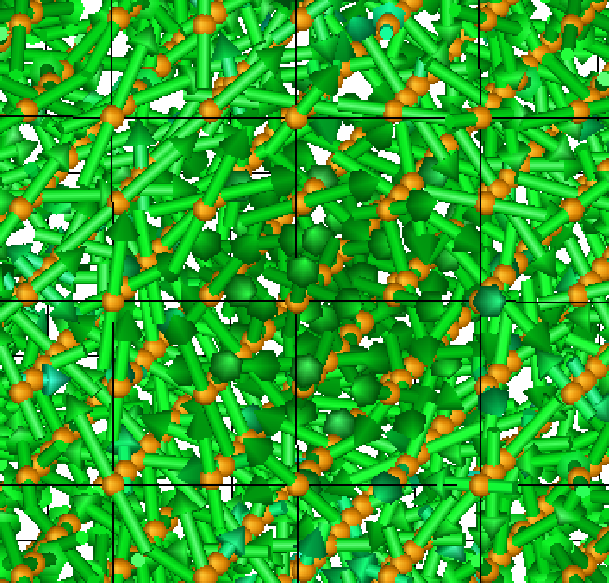} 
\vspace{-0.0cm} \\
(b)
\vspace{5mm} \\
\end{center}
\end{minipage}
\end{center}
\vspace{0cm}
\caption{(a) Magnetic structure of the vortex-type skyrmion MHSDW for $q=0.3$.
(b) Enlarged view showing a vortex spin structure.
}
\label{fig-skvor0.3}
\end{figure}

\begin{figure}[htbp]
\begin{center}
\includegraphics[width=8cm]{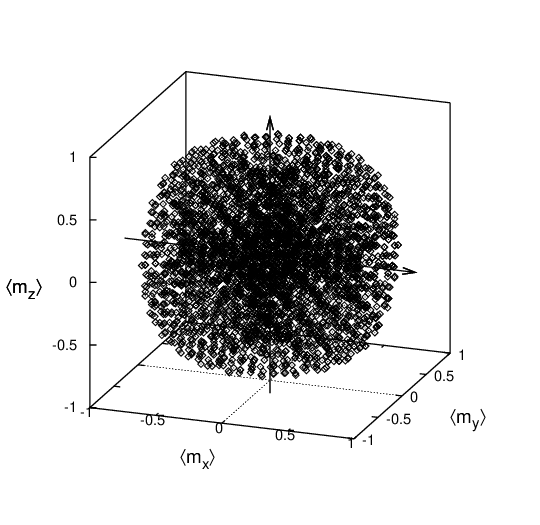}
\end{center}
\vspace{0cm}
\caption{Local moment distribution for the vortex-type skyrmion MHSDW ($q=0.3$).
}
\label{fig-skvorlm}
\end{figure}

\begin{figure}[htbp]
\begin{center}
\includegraphics[width=8cm]{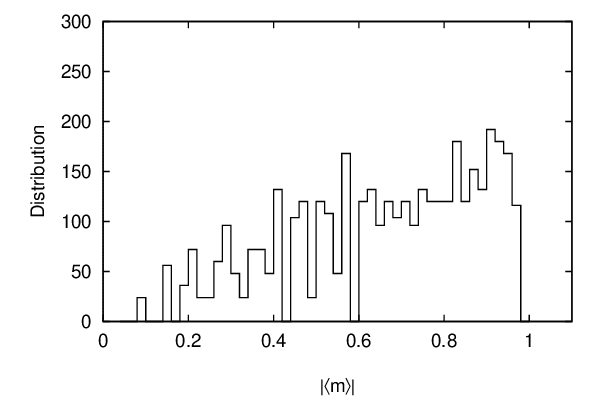}
\end{center}
\vspace{0cm}
\caption{Amplitude distribution of the local magnetic moments for the vortex-type skyrmion MHSDW ($q=0.3$).  The distribution was obtained from 4000 local moments in the $10 \times 10 \times 10$ fcc unit cells.
}
\label{fig-skvorabslm}
\end{figure}

The wave vectors $\bQ_{n}$ in $\gamma$-Fe type MHSDW discussed in the last subsection are neither perpendicular nor parallel to their helical rotational planes.  We consider here the case that the wave vectors $\bQ_{n}$ are perpendicular to their rotational planes: 
$\bQ_{1} = (q, 0, 0) 2\pi/a$, $\bQ_{2} = (0, q, 0) 2\pi/a$, $\bQ_{3} = (0, 0, q) 2\pi/a$.
The AVS image of the magnetic structure for $q=0.3$ is presented in Fig. \ref{fig-skvor0.3}.
We observe a particle-like magnetic structure with size $\lambda=a/q$.  
Each particle shows a vortex structure.  Thus, the MHSDW forms a skyrmion lattice with size $\lambda$.  Note that each vortex-like (antivortex-like) particle with size $\lambda$ is surrounded by 8 antivortex-like (vortex-like) particles with the same size because of a relation $\langle \bm_{l+\tau} \rangle = - \langle \bm_{l} \rangle$ for $\boldsymbol{\tau} = (\lambda/2) (\pm 1, \pm 1, \pm 1)$.
The local moment distribution of the skyrmion-type MHSDW is presented in Fig. \ref{fig-skvorlm}.
The distribution is approximately uniform in direction, being similar to that of the $\gamma$-Fe type MHSDW in Fig. \ref{fig-femhsdwlm}.
In the case of itinerant electron system, the amplitudes of local moments are variable and thus yield a broad distribution as presented in Fig. \ref{fig-skvorabslm}.

\begin{figure}[htbp]
\begin{center}
\includegraphics[width=12cm]{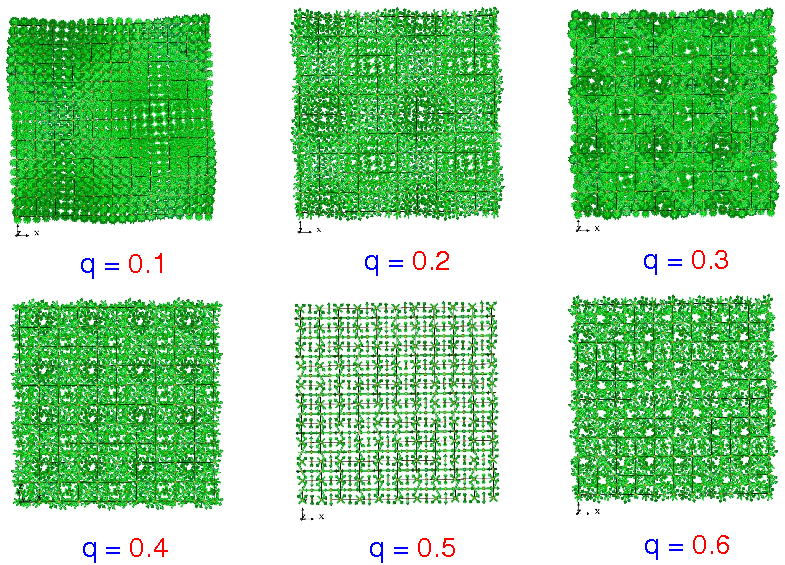}
\end{center}
\vspace{0cm}
\caption{ Vortex-type skyrmion structure projected onto the $x$-$y$ plane ($q=0.1 \sim 0.6$).
}
\label{fig-skvorxy1}
\end{figure}

We examined the change of magnetic structure with increasing $q$.  Their structures projected onto the $x$-$y$ plane are presented in Fig. \ref{fig-skvorxy1}.  With increasing $q$, we observe that the particle size shrinks according to the relation $\lambda=a/q$.
When the size $\lambda$ becomes comparable to the lattice parameter $a$, the particle picture, however, disappears as found for $q=0.5$ in Fig. \ref{fig-skvorxy1}.

\begin{figure}[htbp]
\begin{center}
\includegraphics[width=12cm]{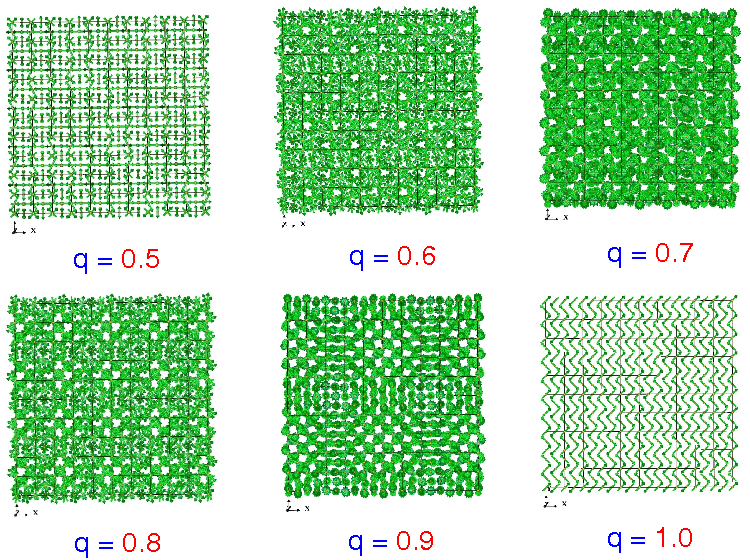}
\end{center}
\vspace{0cm}
\caption{Vortex-type skyrmion structures projected onto the $x$-$y$ plane ($q=0.5 \sim 1.0$).
}
\label{fig-skvorxy2}
\end{figure}

\begin{figure}[htbp]
\begin{center}
\begin{minipage}{9cm}
\begin{center}
\includegraphics[width=9cm]{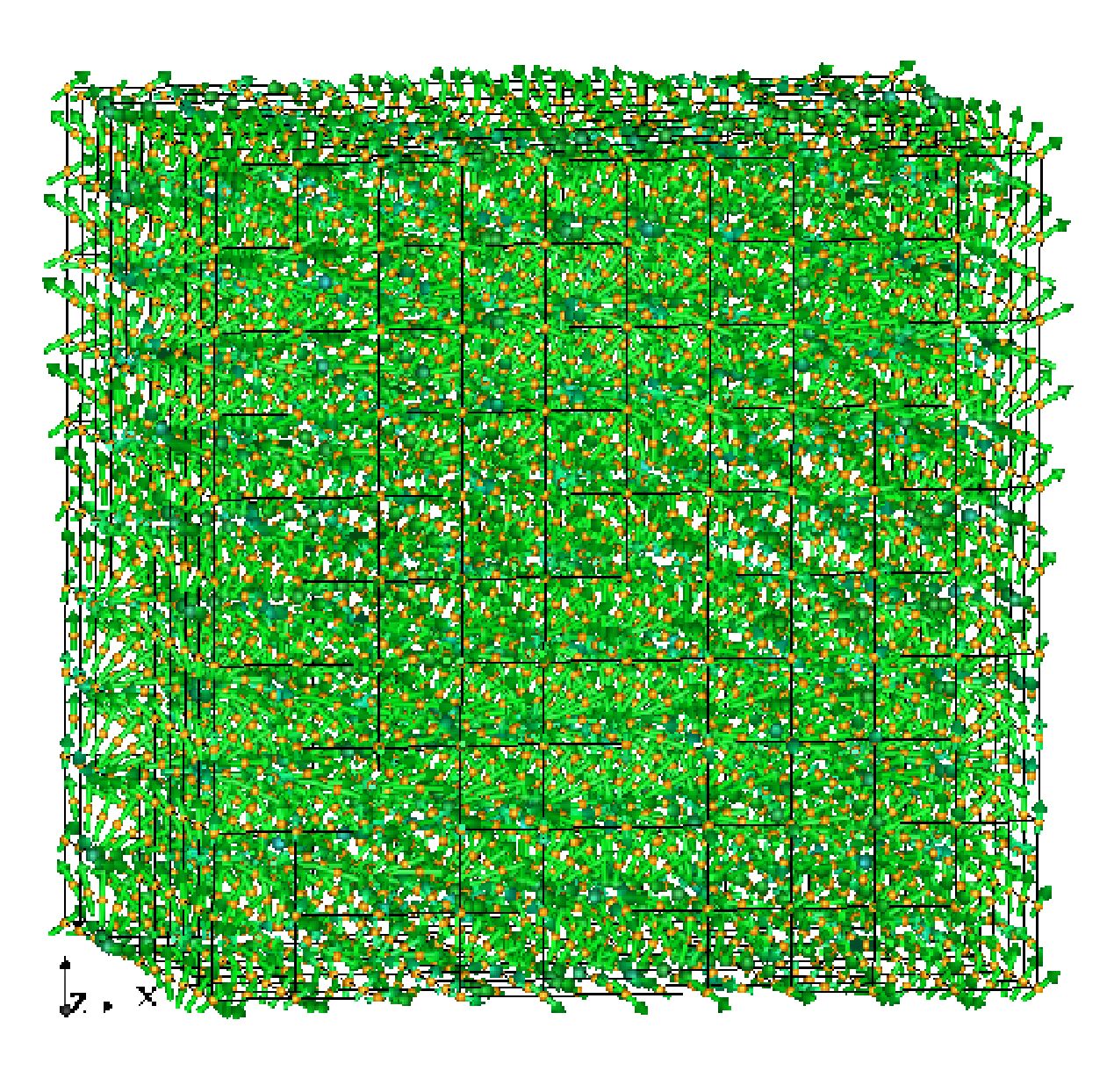}
\vspace{-1cm} \\
(a)
\vspace{5mm} \\
\end{center}
\end{minipage}
\hspace{0.7cm}
\begin{minipage}{5cm}
\begin{center}
\vspace*{3.2cm}
\includegraphics[width=5cm]{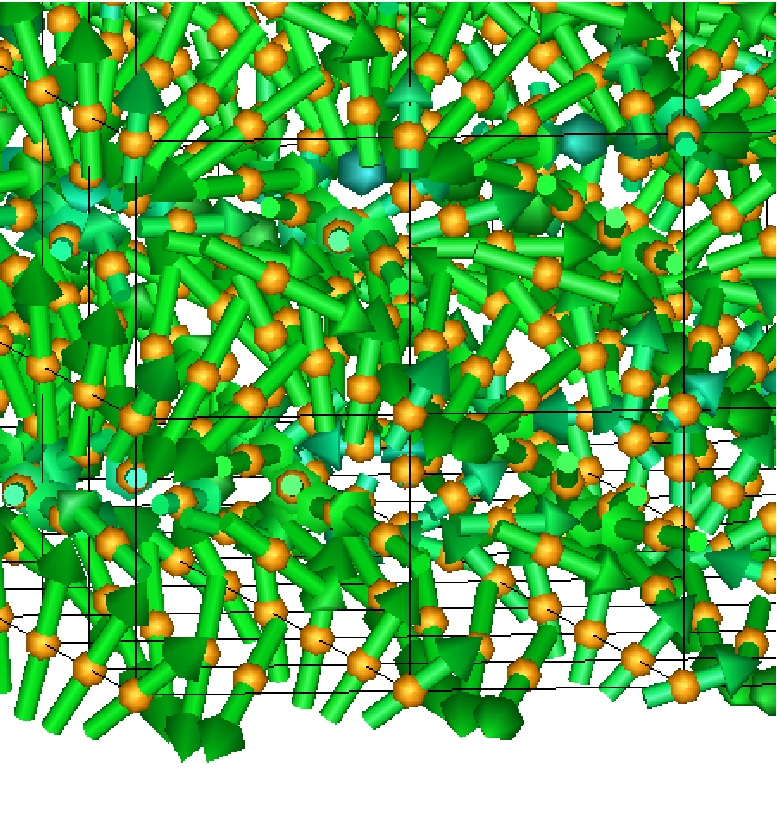} \\
(b)
\vspace{5mm} \\
\end{center}
\end{minipage}
\end{center}
\vspace{0cm}
\caption{ (a) Antiferromagnetic skyrmion structure with size $\tilde{\lambda} = 5a \ (q=0.9)$
(b) Enlarged view showing a local spin structure of the antiferromagnetic skyrmion.
}
\label{fig-skvoraf}
\end{figure}

We have explored further for a new type of magnetic structure
 increasing $q$ from 0.5 to 1.0. Obtained magnetic structures projected onto the $x$-$y$ plane are shown in Fig. \ref{fig-skvorxy2}.  The vortex-like structure is not clear in this range because $\lambda \sim a$, but we observe that a new particle structure with size $\tilde{\lambda} = a/2(1-q)$ emerges when $q \rightarrow 1$, as seen in the case of $q=0.9$.
This is an AF-base skyrmion, though the particle nature is not so clear in the 3 dimensional AVS image (see Fig. \ref{fig-skvoraf}).   To clarify this picture, let us consider first the magnetic structure for $q = 1$ consisting of 3Q antiferromagnetic MHSDW states.
\begin{align}
\langle \bm_{l} \rangle &= m \big( \be_{1} \cos (\hat{\bQ}_{3} \cdot \bR_{l}) +
\be_{2} \cos (\hat{\bQ}_{1} \cdot \bR_{l}) + \be_{3} \cos (\hat{\bQ}_{2} \cdot \bR_{l}) \big) ,
\label{3qmlq1}
\end{align}
where $\hat{\bQ}_{1} = (1, 0, 0) 2\pi/a$, $\hat{\bQ}_{2} = (0, 1, 0) 2\pi/a$, 
$\hat{\bQ}_{3} = (0, 0, 1) 2\pi/a$.
When $\bQ_{n} \neq \hat{\bQ}_{n}$, we can express Eq. (\ref{3qm}) as follows, defining $\tilde{\bq}_{n}$ by means of $\tilde{\bq}_{n} = \hat{\bQ}_{n} - \bQ_{n}$.
\begin{align}
\langle \bm_{l} \rangle &= m_{{\rm AF}1}(\bR_{l}) \big( \be_{2} \cos (\tilde{\bq}_{1} \cdot \bR_{l} - \alpha_{1}) -
\be_{3} \sin (\tilde{\bq}_{1} \cdot \bR_{l} - \alpha_{1}) \big)  \nonumber \\
&+
m_{{\rm AF}2}(\bR_{l}) \big( \be_{3} \cos (\tilde{\bq}_{2} \cdot \bR_{l} - \alpha_{2}) -
\be_{1} \sin (\tilde{\bq}_{2} \cdot \bR_{l} - \alpha_{2}) \big) \nonumber \\
&+
m_{{\rm AF}3}(\bR_{l}) \big( \be_{1} \cos (\tilde{\bq}_{3} \cdot \bR_{l} - \alpha_{3}) -
\be_{2} \sin (\tilde{\bq}_{3} \cdot \bR_{l} - \alpha_{3}) \big) .
\label{3qmtilde}
\end{align}
Here $m_{{\rm AF}n}(\bR_{l}) = m \cos (\hat{\bQ}_{n} \cdot \bR_{l})$.

Equation (\ref{3qmtilde}) indicates that the MHSDW can be regarded as a superposition of the  `helical' SDW with the wave vectors $\tilde{\bq}_{n}$ whose amplitudes of local moments change their sign antiferromagnetically.  Thus, one can regard the MHSDW as an AF skyrmion structure.  When $\tilde{q} = 1-q$ is small as in the case of $q=0.9$ in Fig. \ref{fig-skvorxy2}, the AF skyrmion picture may be suitable.  It should be noted that the AF skyrmion structure is not vortex-like any more and has a size $\tilde{\lambda} = a/2\tilde{q}$, because the amplitudes of local moments change alternatively their direction according to $m_{{\rm AF}n}(\bR_{l})$.

\subsection{Hedgehog-type skyrmion MHSDW}

\begin{figure}[htbp]
\begin{center}
\begin{minipage}{9cm}
\begin{center}
\includegraphics[width=9cm]{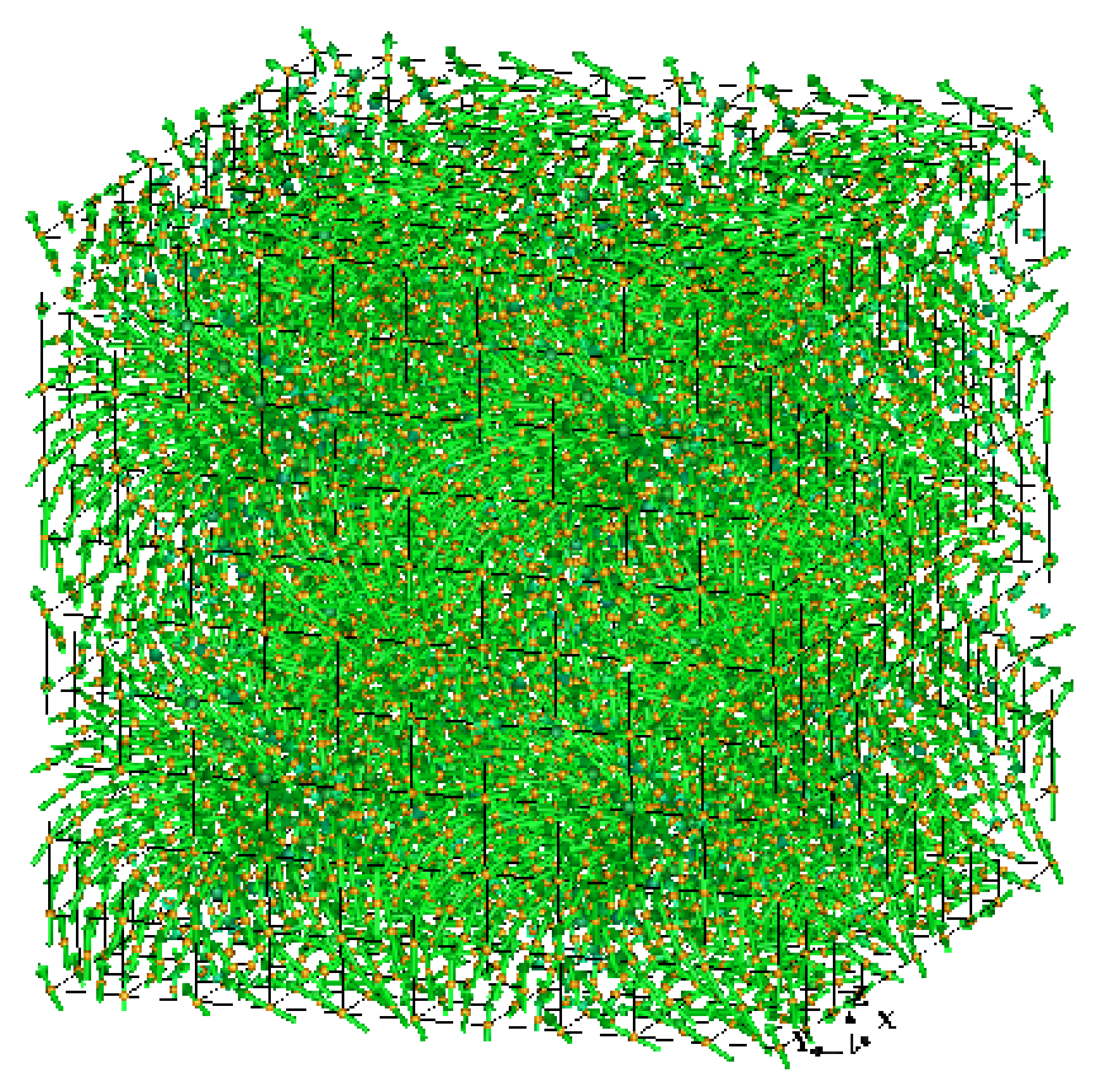}
\vspace{-0.5cm} \\
(a)
\vspace{5mm} \\
\end{center}
\end{minipage}
\hspace{0.7cm}
\begin{minipage}{5.2cm}
\begin{center}
\vspace*{3cm}
\includegraphics[width=5.2cm]{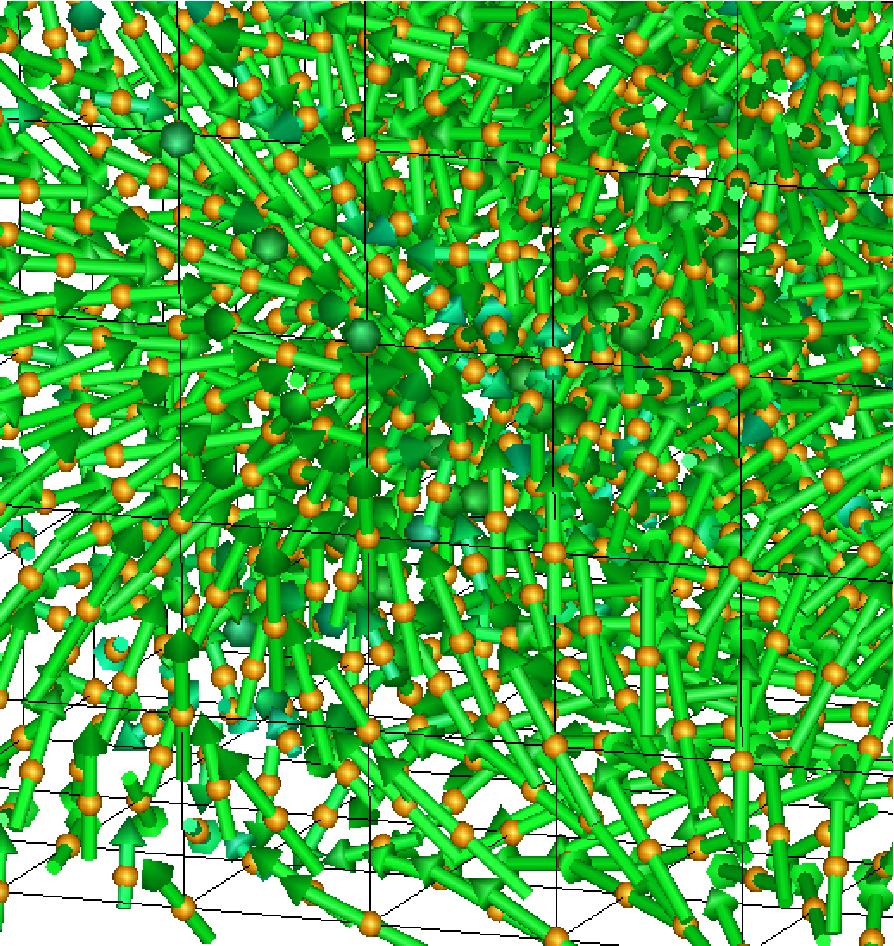} 
\vspace{-0.0cm} \\
(b)
\vspace{5mm} \\
\end{center}
\end{minipage}
\end{center}
\vspace{0cm}
\caption{ (a) Magnetic structure of the hedgehog-type skyrmion MHSDW for $q=0.2$.
(b) Enlarged view showing a local spin structure of the hedgehog-type skyrmion.
}
\label{fig-skhghg}
\end{figure}

We also considered the case that the wave vectors $\bQ_{n}$ are parallel to their rotational planes: $\bQ_{1} = (0, 0, q) 2\pi/a$, $\bQ_{2} = (q, 0, 0) 2\pi/a$, $\bQ_{3} = (0, q, 0) 2\pi/a$.
As shown in Fig. \ref{fig-skhghg}, we observe a hedgehog-like particle structure with size $\lambda = a/q$.  This structure is different from the vortex-type skyrmions discussed in the last subsection, though we find the same local-moment and amplitude distributions.
\begin{figure}[htbp]
\begin{center}
\includegraphics[width=12cm]{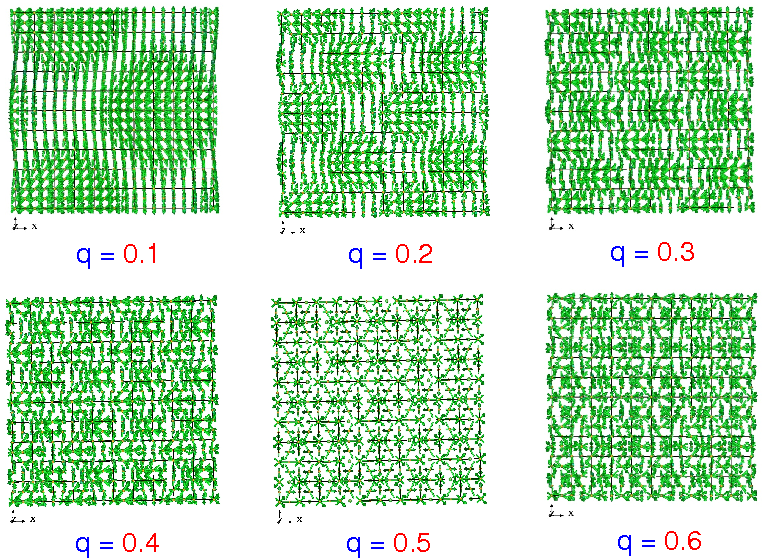}
\end{center}
\vspace{0cm}
\caption{Hedgehog-type skyrmion structures projected onto the $x$-$y$ plane ($q=0.1 \sim 0.6$).
}
\label{fig-skhghgxy}
\end{figure}

With increasing $q$ value, the size of particles decreases according to the relation $\lambda = a/q$ as seen in Fig. \ref{fig-skhghgxy}, and a particle picture disappears again for $q=0.5$.  We have also an AF-base hedgehog skyrmion structure for small $\tilde{q} = 1 - q$ as discussed in the last subsection.  The hedgehog-type MHSDW for $q=1$ agrees with the $\gamma$-Fe type MHSDW with $q=0$.  Finally we remark that the hedgehog-type and skyrmion-type MHSDW are degenerate in the lowest order GL theory presented in \S 2 because the Landau coefficients in both structures agree with each other due to high crystal symmetry of the fcc lattice.

\section{Generalized Hartree-Fock Calculations for Magnetic Structure on the FCC Lattice}

We have  demonstrated on the basis of the GL theory and the AVS image method that the 3Q-MHSDW are possible in the itinerant electron system, and they include the magnetic skyrmion structures as well as the $\gamma$-Fe type MHSDW.  It is not clear, however, whether or not such MHSDW are  possible from a microscopic point of view.  In this section, we present some numerical results of microscopic calculations suggesting the existence of a skyrmion-type MHSDW.

The Hubbard model~\cite{hub63,hub64} is the standard model describing itinerant electron magnetism~\cite{moriya85,kake13}.  We start from the single-band Hubbard model on the fcc lattice as follows.
\begin{eqnarray}
H = \sum_{i,\sigma} \epsilon_{0} n_{i\sigma} 
+ \sum_{i, j, \sigma} t_{i j} a_{i \sigma}^{\dagger} a_{j \sigma}
+ \sum_{i} U n_{i \uparrow}n_{i \downarrow} \ .
\label{hub-h}
\end{eqnarray}
Here $\epsilon_{0}$ is the atomic level,  $t_{i j}$ is the nearest-neighbor transfer integral between sites $i$ and $j$.  $U$ is the intra-atomic Coulomb interaction energy parameter.  
$a_{i \sigma}^{\dagger} (a_{i \sigma})$ is the creation 
(annihilation) operator for an electron with spin $\sigma$ on site $i$,
and $n_{i\sigma}=a_{i \sigma}^{\dagger} a_{i \sigma}$ is the number
operator on site $i$ for spin $\sigma$. 

In order to deal with the noncollinear magnetic structure of the MHSDW, we adopt here the generalized Hartree-Fock approximation.  In this approximation, we introduce locally rotated coordinates on each site, and adopt on each site the Hartree-Fock approximation.  Then, we obtain the following Hamiltonian.
\begin{eqnarray}
H = \sum_{i \alpha j \gamma} a_{i \alpha}^{\dagger} H_{i \alpha j \gamma} a_{j \gamma}
- \sum_{i} \frac{1}{4} U ( \langle n_{i} \rangle^2 - \langle \boldsymbol{m}_{i} \rangle^2 ) \ ,
\label{ghf-h}
\end{eqnarray}
\begin{eqnarray}
H_{i \alpha j \gamma} = 
\big[ ( \epsilon_{0} + \frac{1}{2} U \langle n_{i} \rangle ) \delta_{\alpha \gamma} - 
\frac{1}{2} U \langle \boldsymbol{m}_{i} \rangle \cdot (\boldsymbol{\sigma})_{\alpha \gamma}
\big] \delta_{ij} + t_{ij} \delta_{\alpha\gamma} (1-\delta_{ij}) .
\label{hmatrix}
\end{eqnarray}
Here $\boldsymbol{\sigma}$ are the Pauli spin matrices.  $\langle n_{i} \rangle$ and 
$\langle \boldsymbol{m}_{i} \rangle$ are the average local charge and magnetic moment on site $i$, respectively.  They are given by the on-site one electron Green function defined by 
$G_{i\alpha i\gamma}(z) = [(z-H)^{-1}]_{i\alpha i\gamma}$ and the Fermi distribution function $f(\epsilon)$ as follows.
\begin{eqnarray}
\langle n_{i} \rangle = \int d\epsilon f(\epsilon) \frac{(-)}{\pi} {\rm Im} \sum_{\alpha}
G_{i\alpha i\alpha}(z) ,
\label{ghfni}
\end{eqnarray}
\begin{eqnarray}
\langle \boldsymbol{m}_{i} \rangle = \int d\epsilon f(\epsilon) \frac{(-)}{\pi} {\rm Im} \sum_{\alpha}
(\boldsymbol{\sigma} G)_{i\alpha i\alpha}(z) ,
\label{ghfmi}
\end{eqnarray}
where $z = \epsilon + i\delta$, $\delta$ being a positive definite infinitesimal number.
\begin{figure}[htbp]
\begin{center}
\includegraphics[width=9cm]{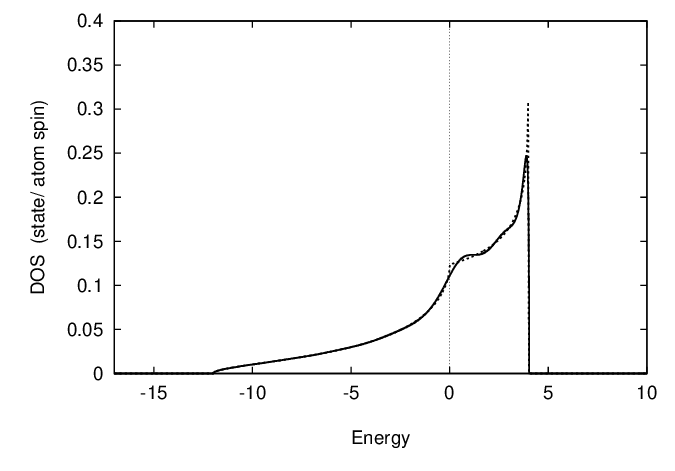}
\end{center}
\vspace{0cm}
\caption{ The fcc density of states (DOS) for noninteracting Hamiltonian. Solid curve: DOS calculated by the recursion method with the recursion level 10.  Dotted curve: DOS obtained by the Brillouin zone integration.  The energy is measured in unit of $|t|$, $t$ being the nearest-neighbor transfer integral.
}
\label{fig-fccdos}
\end{figure}

We calculated the Green function using the recursion method~\cite{hay75,heine80}.
In this method, we make a unitary transformation of Hamiltonian (\ref{hmatrix}) into a tridiagonal matrix by means of the Lanczos method.  The diagonal Green function $G_{i\alpha i\alpha}(z)$, for example, is then expressed by a continued fraction as follows.
\begin{eqnarray}
G_{i\alpha i\alpha}(z) = 
\cfrac{1}{z - a_{1} - 
\cfrac{|b_{1}|^{2}}
{z - a_{2} -
\cfrac{|b_{2}|^{2}} 
{\ldots \cfrac{\ddots}
{\ldots
- \cfrac{|b_{n-1}|^{2}}
{z - a_{n} - T_{n}(z)} 
}}}} \ .
\label{gcont}
\end{eqnarray}
The recursion coefficients $\{ a_{m}, b_{m} \}$ up to $n$ are obtained recursively from the Hamiltonian matrix elements of a cluster with size $n \times n \times n$ in unit of a lattice constant when we adopt the nearest-neighbor transfer integral $t_{ij}$.  For the coefficients larger than $n$, we adopt their asymptotic values $a_{\infty}$, $b_{\infty}$, so that we obtain an approximate form of the terminator $T_{n}(z)$ as follows.
\begin{eqnarray}
T_{n} \approx T_{\infty} = \frac{1}{2} \Big(
z - a_{\infty} - \sqrt{(z - a_{\infty})^{2} - 4 |b_{\infty}|^{2}}
\Big) \ .
\label{tinfty}
\end{eqnarray}
Similary, we can also obtain the off-diagonal Green functions in spin space (see Appendix G in Ref. (28) for more details).

In the present calculation, we made a large cubic cluster consisting of $10 \times 10 \times 10$ fcc unit cells on a computer, which is further surrounded by 27 cubic clusters with the same size, in order to obtain the recursion coefficients.  
We obtained numerically the coefficients up to the 10th level for each site, and used extrapolated values for higher-order coefficients.  Figure \ref{fig-fccdos} shows the calculated density of states (DOS) for noninteracting Hamiltonian.  The DOS reproduces well that was obtained by the Brillouin zone integration except the Van Hove singularity behavior at the origin.
\begin{figure}[htbp]
\begin{center}
\includegraphics[width=11cm]{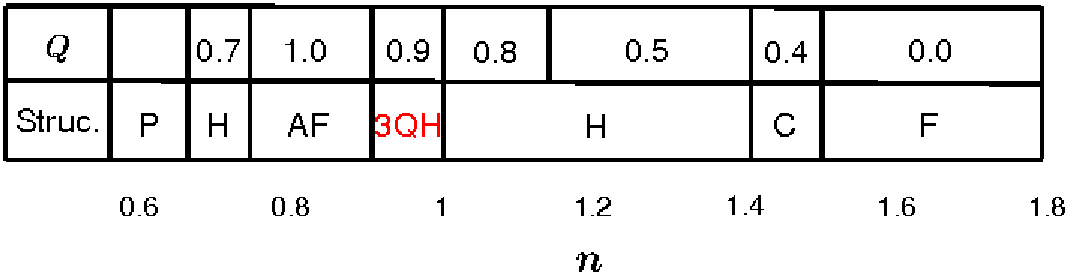}
\end{center}
\vspace{0cm}
\caption{ Magnetic phase diagram for $U/|t| = 8$ as a function of electron number $n$.  F: ferromagnetic structure, C: conical structure, H: helical structure, 3QH: 3Q MHSDW structure, AF: antiferromagnetic structure, and P: paramagnetic state.  Note that the 3QH region ($0.98 < n < 1.00$) is artificially expanded to fill in the phase name.  The magnitude of wave vector $Q$ is expressed in unit of $2\pi/a$.
}
\label{fig-ghfpd}
\end{figure}
\begin{figure}[htbp]
\begin{center}
\includegraphics[width=8cm]{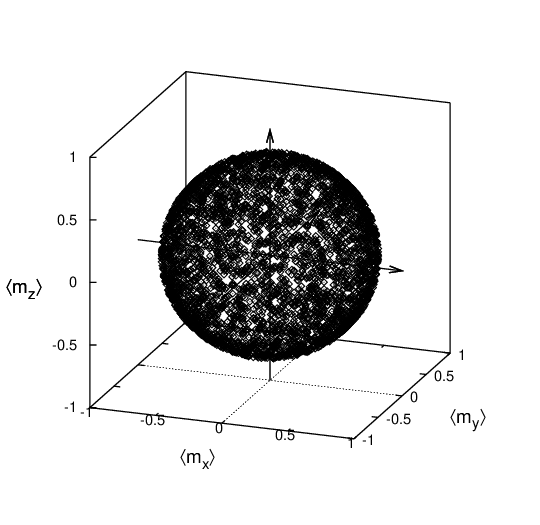}
\end{center}
\vspace{0cm}
\caption{ Local moment distribution for 3QH at $n=0.99$.
}
\label{fig-ghfaflm}
\end{figure}
\begin{figure}[htbp]
\begin{center}
\includegraphics[width=7cm]{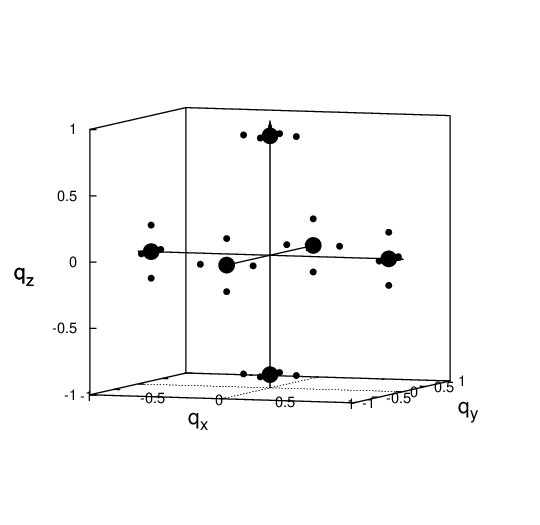}
\end{center}
\vspace{0cm}
\caption{ The $\bQ$ points of the principal Fourier contributions (big spheres) and the second principal contributions (small spheres) in the fcc Brillouin zone.
}
\label{fig-ghffourier}
\end{figure}

In the present work, we adopted $U/|t| = 8$, {\it i.e.}, half a band width of the noninteracting fcc system, bearing in mind typical itinerant magnets.  Here $t$ denotes the nearest-neighbor transfer integral $t_{ij}$.
Assuming the periodic boundary condition among the clusters and starting from a given magnetic structure, we solved
the self-consistent equations (\ref{ghfni}) and (\ref{ghfmi}) for 4000 atoms in the central cluster at zero temperature.  
After having solved Eqs. (\ref{ghfni}) and (\ref{ghfmi}), we compared the energy among the self-consistent sets of solutions for the  ferromagnetic (F), conical (C), helical (H), and 3Q-MHSDW (3QH) structures in order to find the ground-state magnetic structure.

We determined the magnetic structure at the ground state for each electron number $n$.  Calculated magnetic phase diagram is shown in Fig. \ref{fig-ghfpd} as a function of $n$.  We did not find any phase separation in the present calculation.  With decreasing $n$ from 2, the magnetic structure changes from F to C to H to 3QH to AF to H again, and to the paramagnetic state P.  Accordingly, the magnitude of the wave vector $|\bQ|$ increases from 0 to 1 in unit of $2\pi/a$.  In particular, we found the 3QH structure in the region $0.98 < n < 1.00$ between the H and AF structures.  Although the lowest-order GL theory does not necessarily well describe behaviors of the Hubbard model in the intermediate coupling regime, the phase diagram in Fig. \ref{fig-3qhpd} seems to support the appearance of the 3QH structure; the appearance of the 3QH may be understood as crossing the phase boundary $B_{QQH}/B_{1Q} =8$ in Fig. \ref{fig-3qhpd} with decreasing electron number $n$.  
\begin{figure}[htbp]
\begin{center}
\begin{minipage}{9cm}
\begin{center}
\includegraphics[width=9cm]{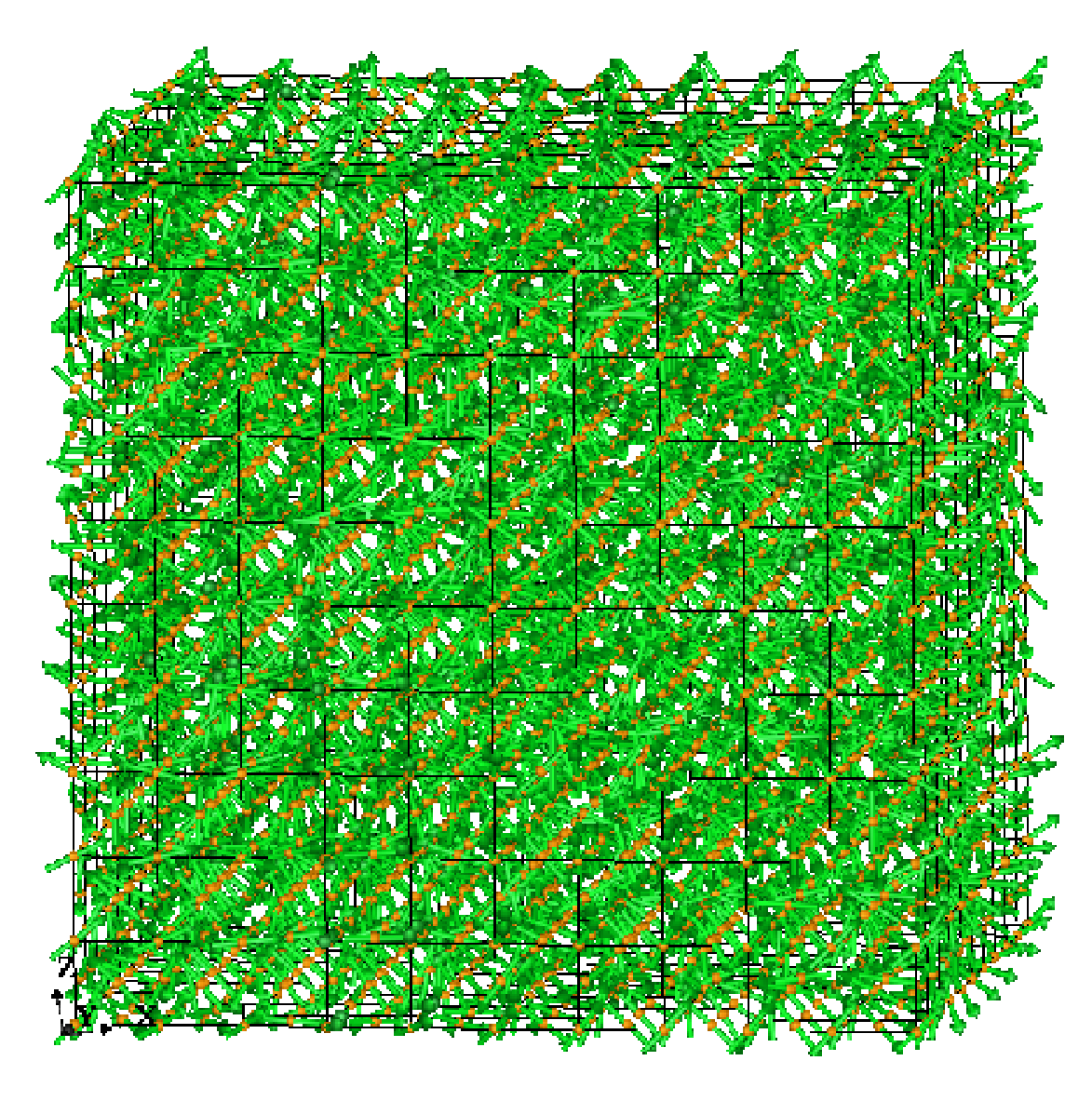}
\vspace{-0.5cm} \\
(a)
\vspace{5mm} \\
\end{center}
\end{minipage}
\hspace{0.7cm}
\begin{minipage}{5cm}
\begin{center}
\vspace*{3.3cm}
\includegraphics[width=5cm]{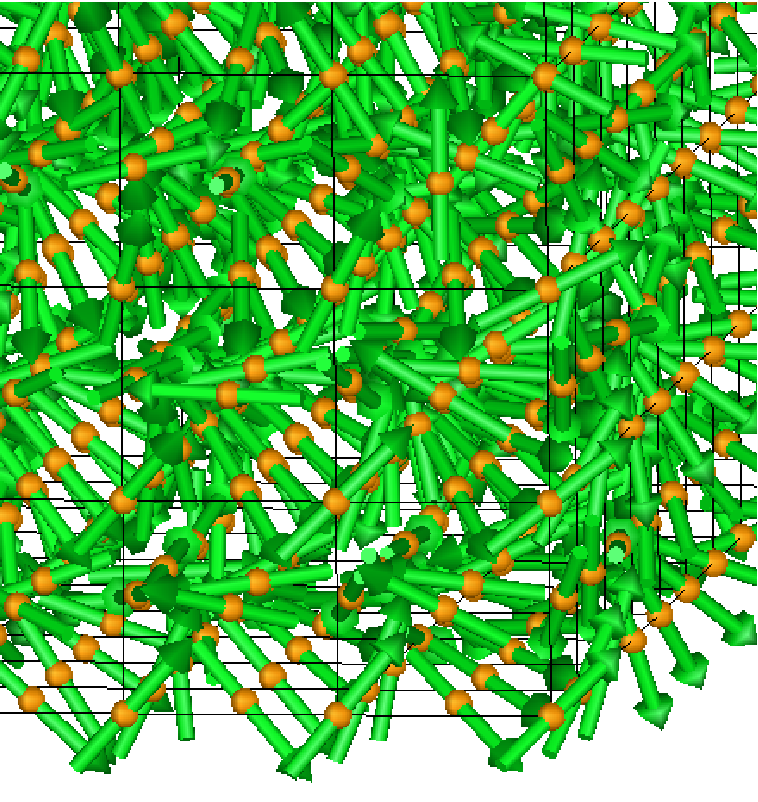} 
\vspace{-0.0cm} \\
(b)
\vspace{5mm} \\
\end{center}
\end{minipage}
\end{center}
\vspace{0cm}
\caption{ (a) AF-base skyrmion structure obtained for $n=0.99$.
(b) Enlarged view showing a local spin structure of AF-base skyrmion.
}
\label{fig-ghfafsk}
\end{figure}
\begin{figure}[htbp]
\begin{center}
\begin{minipage}{9cm}
\begin{center}
\includegraphics[width=9cm]{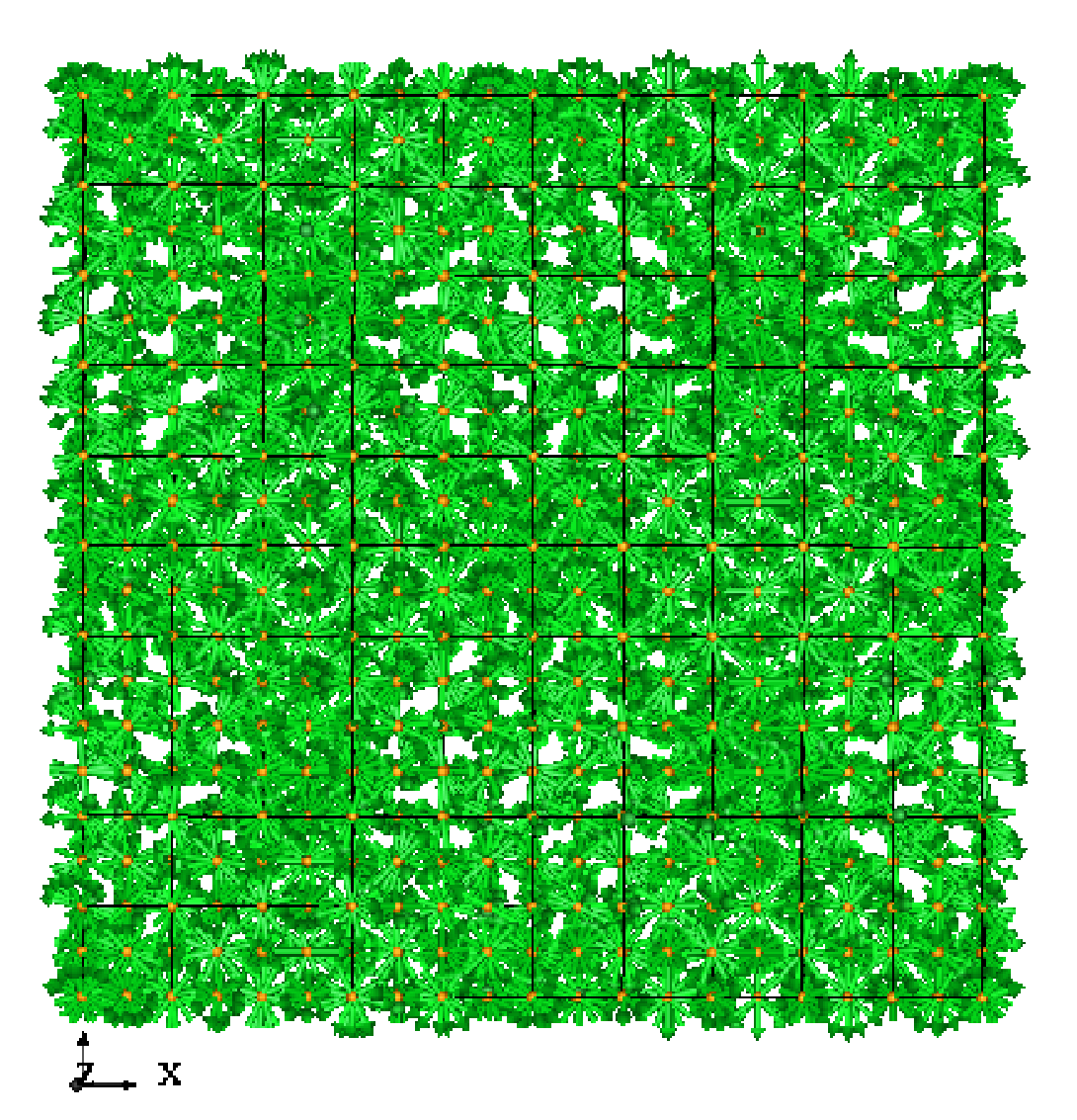}
\vspace{-1cm} \\
(a)
\vspace{5mm} \\
\end{center}
\end{minipage}
\hspace{0.7cm}
\begin{minipage}{5cm}
\begin{center}
\vspace*{3.8cm}
\includegraphics[width=5cm]{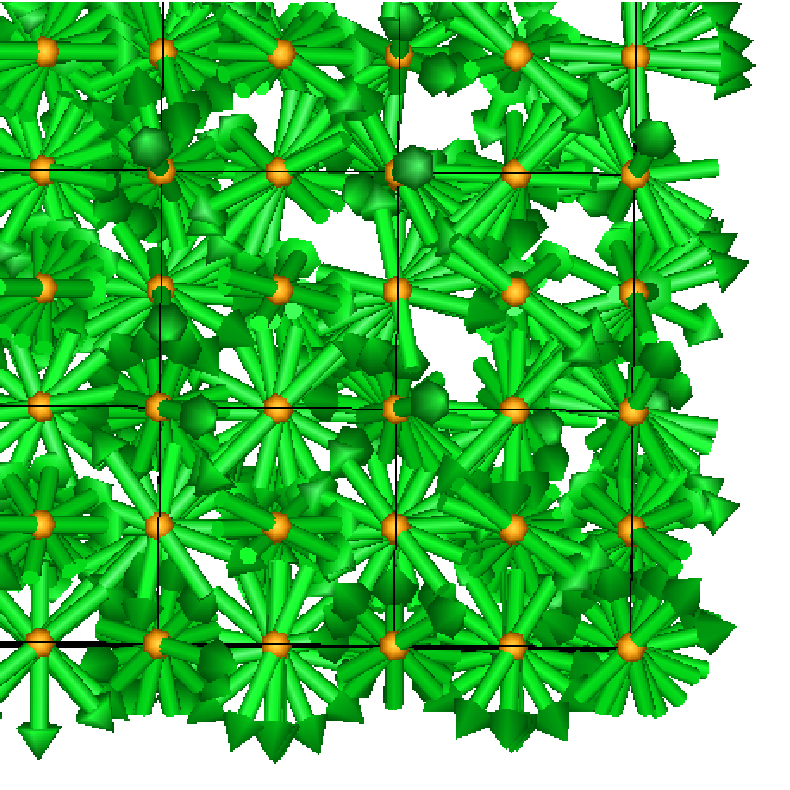} 
\vspace{-0.4cm} \\
(b)
\vspace{5mm} \\
\end{center}
\end{minipage}
\end{center}
\vspace{0cm}
\caption{ (a) AF-base skyrmion structure projected onto the $x$-$y$ plane ($n=0.99$).
(b) Enlarged view showing its local spin configuration.
}
\label{fig-ghfafskxy}
\end{figure}

The local-moment distribution for the 3QH state is shown in Fig. \ref{fig-ghfaflm}.  We find that the distribution is almost uniform in direction.  Calculated 3QH however has no amplitude modulation in contradiction with the GL theory in \S 2; the magnetic moments are distributed on a sphere with radius $|\langle \boldsymbol{m}_{i} \rangle| = 0.77$ in order to reduce the energy loss due to  the Coulomb interaction.  The inconsistency is attributed to the considerably strong Coulomb interaction energy parameter $U/|t|=8$ used in the present calculations; calculated 3QH is no longer regarded as the weak itinerant electron magnet.
(Note that the lowest-order GL theory presented in \S 2 works best for the weak itinerant electron system.)

We also made the Fourier analysis of the magnetic structure using the relation $\bm_{l} = \sum_{\bq} \bm(\bq) \exp (i\bq \cdot \bR_{l})$.  The principal terms with amplitude $|\boldsymbol{m}(\bQ)| = 0.28$ are located at $\bQ = (\pm q, 0, 0)2\pi/a$, $(0, \pm q, 0)2\pi/a$, $(0, 0, \pm q)2\pi/a$ with $q=0.9$, and the second principal  terms with $|\boldsymbol{m}(\bQ)| = 0.05$ are located at 
$\bQ = (\pm q, \pm 0.2, 0)2\pi/a$, $(\pm 0.2, \pm q, 0)2\pi/a$, $(0, \pm 0.2, \pm q)2\pi/a$ as shown in Fig \ref{fig-ghffourier}.  The higher order contributions including the second principal terms are produced so as to reduce the amplitude modulations.  The structure is therefore basically the AF-base skyrmion as discussed in \S 3.2.  Indeed, we find in the AVS image analysis the particle-like units with size $\lambda = 2/(1-q)$ as shown in Figs. \ref{fig-ghfafsk} and \ref{fig-ghfafskxy}, though it is not so clear in the 3 dimensional image.  (Fig. \ref{fig-ghfafsk} should be compared with Fig. \ref{fig-skvoraf}, and Fig. \ref{fig-ghfafskxy} (a) should be compared with the panel of $q=0.9$ in Fig. \ref{fig-skvorxy2}.)
\begin{figure}[htbp]
\begin{center}
\includegraphics[width=9cm]{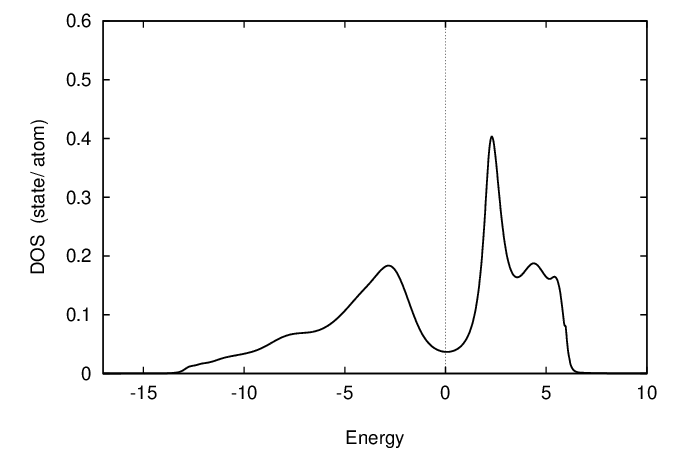}
\end{center}
\vspace{0cm}
\caption{ Calculated density of states (DOS) per atom for $n=0.99$.
}
\label{fig-ghfdos}
\end{figure}

Calculated density of states for the AF-base skyrmion MHSDW structure is presented in Fig. \ref{fig-ghfdos}.  We find a deep valley at the Fermi level as a result of the formation of 3QH state.  This causes a large kinetic energy gain, and thus this is the electronic origin of the stability of the AF skyrmion MHSDW.

\section{Summary and Discussions}

We have investigated the multiple helical spin density waves (MHSDW) and magnetic skyrmions in itinerant electron system on the fcc lattice using the Ginzburg-Landau (GL) theory without the DM interaction and the Application Visualization System (AVS) image method.  The magnetic skyrmion structures are a MHSDW with a particle nature.  We observed using the AVS that the vortex-type skyrmion lattice structures with amplitude modulations appear when the $\bQ$ vectors of the MHSDW are perpendicular to the rotational planes of helical SDW, while the hedgehog-type skyrmion lattice structures appear when the $\bQ$ vectors are parallel to the planes.  We also found using the AVS that the AF-base skyrmion structure appears when the wave vectors approach to the AF ones on the Brillouin zone boundary.

We verified using the GL theory that the magnetic skyrmions mentioned above can be stabilized in the itinerant electron system in a certain region of the space of Landau's coefficients ($-4 < B_{QQH}/B_{1Q} < 8$ and $A_{Q}/B_{1Q} < 0$).  It should be noted that the amplitude modulations of local moments are favorable for the 3Q-MHSDW in the weakly correlated itinerant electron system.  This feature does not exist in the Heisenberg system at the ground state.

We have also examined the possibility of the itinerant electron 3Q-MHSDW using the single-band Hubbard model on the fcc lattice.  For an intermediate strength of Coulomb interaction parameter $U/|t| = 8$, we found an AF-base skyrmion magnetic structure at $n=0.99$, which is located between the helical and AF structures when the electron number $n$ is decreased.
There is no amplitude fluctuation in this case because of a considerably large $U/|t|$ parameter.  A competition between the  long-range ferro- and antiferro-magnetic interactions can stabilize the AF-base magnetic skyrmion near the half-filling regime, and causes a deep valley structure in the density of states near  the Fermi level leading to a large kinetic energy gain.

In the present work, we relied on the AVS image method and numerical calculations for understanding the MHSDW and magnetic skyrmion structures.  Analytic properties of magnetic skyrmions in itinerant electron systems with variable amplitudes of magnetic moment were not clarified in the present paper.  For example, we have not clarified the analytic property of skyrmion number in the itinerant system with variable amplitudes.
Moreover, we restricted ourselves to the fcc lattice in the present work.  We have to investigate the itinerant system without inversion symmetry on order to understand the interplay between the long-range competing magnetic interactions and the DM interaction in the formation of the magnetic skyrmions.

In the microscopic calculations, we verified the possible stability of the AF-base skyrmion only for $U/|t| = 8$.  In order to clarify the stability of the skyrmions, it is desirable to perform the same calculations over all $U/|t|$.  Moreover the present approach based on the generalized Hartree-Fock approximation does not necessarily guarantee the global minimum of the free energy.  Numerical studies with use of the molecular spin dynamics method are highly desired in the future work.

\begin{acknowledgment}

%\acknowledgment

The authors would like to express their thanks to Prof. T. Uchida
 for valuable discussions and comments on the present work. 

\end{acknowledgment}

%\appendix

%%%

\end{document}